\documentstyle[10pt]{article}
\input{tcilatex}

\begin{document}

\title{{\bf \ {\large Comment on Ricci Collineations for spherically symmetric
space-times} }}
\author{P. S. Apostolopoulos$^{*}$ and M. Tsamparlis\thanks{%
Department of Physics, Section of Astronomy-Astrophysics-Mechanics,
University of Athens, ZOGRAFOS 15783, ATHENS, GREECE.
E-mails:papost@cc.uoa.gr, mtsampa@phys.uoa.gr}}
\maketitle

\begin{abstract}
It is shown that the results of the paper by Contreras et al. [Contreras,
G., Nunez, L. A., Percoco, U. {\em Ricci Collineations for Non-degenerate,
Diagonal and Spherically Symmetric Ricci Tensors} (2000). {\it Gen. Rel.
Grav.} {\bf 32}, 285-294] concerning the Ricci Collineations in spherically
symmetric space-times with non-degenerate and diagonal Ricci tensor do not
cover all possible cases. Furthermore the complete algebra of Ricci
Collineations of certain Robertson-Walker metrics of vanishing spatial
curvature are given.
\end{abstract}

The determination of Ricci collineations in a spherically symmetric
space-time is an interesting and difficult problem. The static case has been
considered in a series of papers \cite{Bokhari-Qadir}, \cite
{Amir-Bokhari-Qadir}, \cite{Farid-Qadir-Ziad} however in an incomplete way 
\cite{Bertolotti-Contreras-Nunez-Percoco-Carot}. The non-static case, which
is far more difficult and contains the static case as a special case, has
been considered recently by Contreras et al. \cite{Conteras-Nunez-Percoco}.
The purpose of the present comment is to show by counter examples that this
later approach although correct it is also incomplete.

In \cite{Conteras-Nunez-Percoco} the authors consider the Spherically
Symmetric (SS) metric which in natural coordinates has the form:

\begin{equation}
ds^{2}=-e^{2\nu (t,r)}dt^{2}+e^{2\lambda (t,r)}dr^{2}+Y^{2}(t,r)(d\theta
^{2}+\sin ^{2}\theta d\phi ^{2})  \label{sx1}
\end{equation}
and they look for Ricci Collineations (RCs) $X^{a}$ defined by ${\cal L}_{%
{\bf X}}R_{ab}=0$ where $R_{ab}$ is the Ricci tensor associated with the
metric (\ref{sx1}). To do this they use the following result of \cite
{Carot-Nunez-Percoco}:

{\em The proper RCs of the space-time (\ref{sx1}) whose Ricci tensor is
non-degenerate, are of the form:}

\begin{equation}
{\bf \xi }=\xi ^{t}(t,r)\partial _{t}+\xi ^{r}(t,r)\partial _{r}  \label{sx2}
\end{equation}
and they conclude that ''the form of the most general RC vector is the one
given in (\ref{sx2}) plus linear combinations, with constant coefficients,
of the Killing vectors for spherical symmetry''. Subsequently they compute
64 different classes of SS metrics which admit (not necessarily all proper)
RCs and classify them in Table 1 using the time and the radial first
derivatives of the components of the Ricci tensor.

However it has been shown \cite{Tsamparlis-Apostolopoulos} that the form (%
\ref{sx2}) of the generic RC in SS metrics is not the most general one.
Consequently one should expect that the results of \cite
{Conteras-Nunez-Percoco} are also incomplete. This is indeed the case as we
show by the following two counter examples based on the very examples given
in \cite{Conteras-Nunez-Percoco}.

Consider the RW metric with $k=0$:

\begin{equation}
ds^{2}=-dt^{2}+F^{2}(t)\left( dr^{2}+r^{2}d\theta ^{2}+r^{2}\sin ^{2}\theta
d\phi ^{2}\right) .  \label{sx3}
\end{equation}
The components of the Ricci tensor are:

\begin{equation}
R_{00}=-3\frac{F_{,tt}}{F},\quad R_{11}=\Delta ,\quad R_{22}=r^{2}\Delta
,\quad R_{33}=r^{2}\Delta \sin ^{2}\theta  \label{sx5}
\end{equation}
where:

\begin{equation}
\Delta =FF_{,tt}+2\left( F_{,t}\right) ^{2}  \label{sx6}
\end{equation}
and a comma denotes partial derivative w.r.t the index that follows.
According to the classification given in \cite{Conteras-Nunez-Percoco} the
Ricci tensor (\ref{sx5}) belongs to the family number 7. One distinguishes
two cases according to the constancy of $\Delta $.

\underline{Case $\Delta =const.=\pm a^{2}\neq 0$}

In \cite{Conteras-Nunez-Percoco}\ for this case one finds the RC ${\bf X}%
_1=e^{\epsilon a^2}\left| R_{00}(\tau )\right| ^{-1/2}\partial _t$ (cf eq.
(33)) which has been found previously by Green et al. \cite{Green et al.}.
However it is easy to show that the following vector fields are {\em proper}
RCs:

\begin{equation}
{\bf X}_{2}=ar\cos \phi \sin \theta \partial _{\tilde{\tau}}-\epsilon a^{-1}%
\tilde{\tau}(t)\left\{ \cos \phi \left[ \sin \theta \partial _{r}+\frac{\cos
\theta }{r}\partial _{\theta }\right] -\frac{\sin \phi }{r\sin \theta }%
\partial _{\phi }\right\}  \label{sx8}
\end{equation}

\begin{equation}
{\bf X}_{3}=ar\sin \phi \sin \theta \partial _{\tilde{\tau}}-\epsilon a^{-1}%
\tilde{\tau}(t)\left\{ \sin \phi \left[ \sin \theta \partial _{r}+\frac{\cos
\theta }{r}\partial _{\theta }\right] +\frac{\cos \phi }{r\sin \theta }%
\partial _{\phi }\right\}  \label{sx9}
\end{equation}

\begin{equation}
{\bf X}_{4}=ar\cos \theta \partial _{\tilde{\tau}}-\epsilon a^{-1}\tilde{\tau%
}(t)\left[ \cos \theta \partial _{r}-\frac{\sin \theta }{r}\partial _{\theta
}\right]  \label{sx10}
\end{equation}
where:

\begin{equation}
\tilde{\tau}(t)=\int \left| R_{00}\right| ^{1/2}dt  \label{sx11}
\end{equation}
and $\epsilon =sign(R_{00}\Delta )$.

It is well known that if the Ricci tensor is everywhere of rank 4 (i.e. non
degenerate) the family of (at least $C^{2}$) RCs is a Lie algebra of smooth
vector fields of maximum dimension $10$ \cite{Hall-Roy-Vaz}. Hence the 6 KVs
of the RW space-time plus the 4 RCs ${\bf X}_{1},{\bf X}_{2},{\bf X}_{3},%
{\bf X}_{4}$ above generate the {\em complete} Lie algebra of RCs of this
case.

\underline{Case $\Delta _{,t}\neq 0$}

The metric (\ref{sx3}) with $F(t)=(M/2b)u^{2}$ and $u=(6t/M)^{1/3}$ where $M$
and $b$ are constants \cite{Chan et al.} is a second example in \cite
{Conteras-Nunez-Percoco}. In this case $\Delta _{,t}\neq 0$. According to
the classification given in \cite{Conteras-Nunez-Percoco} this space-time
admit the RC:

\begin{equation}
{\bf \xi }=\sqrt{\frac{3}{2}}t\partial _{t}+\frac{r}{\sqrt{6}}\partial _{r}.
\label{sx12}
\end{equation}
which, as it is correctly stated, is not a proper RC but a HVF . Again it
can be shown that the following {\em proper} RCs complete the Lie algebra of
RCs for this space-time:

\begin{equation}
{\bf X}_{1}=6rt\cos \phi \sin \theta \partial _{t}+\frac{\left[
r^{2}M^{2/3}-(6t)^{2/3}b^{2}\right] \cos \phi \sin \theta }{M^{2/3}}\partial
_{r}-\frac{\left[ r^{2}M^{2/3}+(6t)^{2/3}b^{2}\right] }{M^{2/3}r}\left[ \cos
\phi \cos \theta \partial _{\theta }-\frac{\sin \phi }{\sin \theta }\partial
_{\phi }\right]  \label{sx13}
\end{equation}
\begin{equation}
{\bf X}_{2}=6rt\sin \phi \sin \theta \partial _{t}+\frac{\left[
r^{2}M^{2/3}-(6t)^{2/3}b^{2}\right] \sin \phi \sin \theta }{M^{2/3}}\partial
_{r}-\frac{\left[ r^{2}M^{2/3}+(6t)^{2/3}b^{2}\right] }{M^{2/3}r}\left[ \sin
\phi \cos \theta \partial _{\theta }+\frac{\cos \phi }{\sin \theta }\partial
_{\phi }\right]  \label{sx14}
\end{equation}

\begin{equation}
{\bf X}_{3}=6rt\cos \theta \partial _{t}+\frac{\left[
r^{2}M^{2/3}-(6t)^{2/3}b^{2}\right] \cos \theta }{M^{2/3}}\partial _{r}+%
\frac{\left[ r^{2}M^{2/3}+(6t)^{2/3}b^{2}\right] \sin \theta }{M^{2/3}r}%
\partial _{\theta }.  \label{sx15}
\end{equation}
From the above we draw the following conclusions:

\begin{enumerate}
\item  The generic form of the RCs of the SS metric (\ref{sx1}) for the case
where the Ricci tensor is {\em non degenerate and diagonal} is: 
\begin{equation}
{\bf X}=X^{t}(t,r,\theta ,\phi )\partial _{t}+X^{r}(t,r,\theta ,\phi
)\partial _{r}+X^{\theta }(t,r,\theta ,\phi )\partial _{\theta }+X^{\phi
}(t,r,\theta ,\phi )\partial _{\phi }.  \label{sx16}
\end{equation}

\item  The results in \cite{Conteras-Nunez-Percoco}\ do not cover all
possible cases and the question of determining all RCs of a SS metric (\ref
{sx1}) is still open. However this conclusion in no way diminishes the
essential contribution of \cite{Conteras-Nunez-Percoco}.
\end{enumerate}

A complete study of the RCs in RW space-times is under preparation.

\end{document}